\documentclass[prb,twocolumn,showpacs,amsmath,amssymb,preprintnumbers]{revtex4}
\usepackage{dcolumn}
\usepackage{bm}
\usepackage{graphicx}
\usepackage{color}

\begin{document}

\title{Nonmagnetic substitution in pyrochlore iridate Y$_2$(Ir$_{1-x}$Ti$_{x}$)$_2$O$_7$: Structure, magnetism and electronic properties}

\author{Harish Kumar}\affiliation{School of Physical Sciences, Jawaharlal Nehru University, New Delhi - 110067, India.}
\author{A. K. Pramanik}\email{akpramanik@mail.jnu.ac.in}\affiliation{School of Physical Sciences, Jawaharlal Nehru University, New Delhi - 110067, India.}

\begin{abstract}
Tuning of spin-orbit coupling and electron correlation effects in pyrochlore iridates is considered for many interesting phenomena. We have investigated the temperature evolution of structural, magnetic and electronic properties in doped Y$_2$(Ir$_{1-x}$Ti$_{x}$)$_2$O$_7$ ($x$ = 0.0, 0.02, 0.05, 0.10 and 0.15) where the substitution of nonmagnetic Ti$^{4+}$ (3$d^0$) for Ir$^{4+}$ (5$d^5$) amounts to dilution of magnetic network and tuning of these parameters in opposite way. The system retains its original structural symmetry but local structural parameters show an evolution with Ti content. While the magnetic transition temperature is not largely influenced, both magnetic moment and magnetic frustration decreases with Ti doping. Magnetic relaxation measurement shows the parent compound Y$_2$Ir$_2$O$_7$ as well as its Ti doped analogues are in nonequilibrium magnetic state where the magnetic relaxation rate increases with Ti. Temperature dependent Raman measurements indicate no changes in structural symmetry, however, across the magnetic transition temperature an anomaly in A$_{1g}$ Raman mode is observed. Temperature dependent x-ray diffraction data also support the Raman spectroscopy data, however, an evolution of lattice parameters with temperature is observed. The electrical resistivity data of Y$_2$(Ir$_{1-x}$Ti$_{x}$)$_2$O$_7$ series exhibits insulating behavior throughout the temperature range, however, the resistivity decreases with Ti doping. The nature of charge conduction is found to follow power-law behavior in whole series but the validity of this model varies with temperature. A negative magnetoresistance has been observed at low temperature in present series which is explained with weak localized mechanism. Similar to other Ir based oxides, a crossover from negative to positive MR has been observed in present system.
\end{abstract}


\maketitle
\section {Introduction}
In recent times, physics driven interplay between spin-orbit coupling (SOC) and electronic correlation ($U$) has generated lot of interest as it promises to host many topological phases of mater.\cite{pesin,william} The 5$d$ based iridium oxides are of particular interest as SOC effect becomes very prominent in these materials ($\sim$ 0.5 eV) due to heavy character of Ir. Moreover, with relatively weaker $U$ in 5$d$ orbitals, these materials exhibit a comparable scale between relevant energies such as, SOC, $U$ and crystal field effect. This offers an ideal playground to study the fascinating physics which comes out as an interplay between these competing parameters. Pyrochlore based structures (A$_2$Ir$_2$O$_7$ where A = trivalent rare earth elements) further add interest due to its structural arrangement where the A and Ir atoms form corner shared interpenetrating tetrahedra which introduces geometrical frustration  and resultantly new physics.\cite{gingras,yoshii,gardner1,nakatsuji,bramwell,fukazawa1} Theoretical calculation further suggests that inclusion of Dzyaloshinskii-Moriya (DM) type interaction in original nearest-neighbor Heisenberg model which basically generates frustration and hiders magnetic transition at low temperature, induces a long-range ordered magnetic transition at low temperature.\cite{elhajal} This is important at the backdrop of pyrochlore iridates given that these materials simultaneously possess geometrical frustration and reasonable SOC where the later acts as an ingredient for DM interaction. Pyrochlore iridates mostly exhibit antiferromagnetic (AFM) insulating phase at low temperature, however, their physical properties notably evolve with A-site element: from magnetic insulating to paramagnetic (PM) metallic phase.\cite{Matsuhira} 

The Y$_2$Ir$_2$O$_7$ is an interesting compound among pyrochlore iridates because it has nonmagnetic Y$^{3+}$ at A-site which excludes the possibility of \textit{f-d} exchange interactions.\cite{chen} In this sense, the magnetic properties are mostly governed by the Ir-sublattice. The Y$_2$Ir$_2$O$_7$ is a strong insulator and exhibits a bifurcation between zero field cooled and field cooled magnetization data at $\sim$ 160 K.\cite{shapiro,disseler,taira,fukazawa,soda,zhu,harish} There are, however, contradictory data about the low temperature magnetic state in Y$_2$Ir$_2$O$_7$. The muon spin relaxation results show a long-range ordered magnetic state while the neutron diffraction experiments apparently find no sign of magnetically ordered state within the accuracy of instrument.\cite{disseler,shapiro} In another study, weak ferromagnetism (FM) with AFM background is reported for Y$_2$Ir$_2$O$_7$.\cite{zhu} Recently, using magnetic relaxation measurements a nonequilibrium magnetic state in Y$_2$Ir$_2$O$_7$ and its Ru doped samples is reported at low temperature.\cite{harish,kumar} Theoretical calculation has predicted Y$_2$Ir$_2$O$_7$ a possible candidate for novel Weyl semimetal with all-in/all-out (AIAO) type AFM spin order.\cite{wan} Further band calculations employing DFT+DMFT, LDA+DMFT and LSDA+$U$ techniques have shown a transition to an AIAO-type AF state with varying $U$, however, the threshold value of $U$ is not uniquely defined.\cite{shinaoka,Ishii,Hon} Here to be noted that photoemission spectroscopy data suggest $U$ $\sim$ 4 eV in Y$_2$Ir$_2$O$_7$ which is surprisingly high for 5$d$ electron systems.\cite{kalo} Therefore, Y$_2$Ir$_2$O$_7$ definitely comes out as good model system to tune the relative strength of SOC and $U$ and study its interplay.

With an aim to study the effect of site dilution as well as to tune the both SOC and $U$, we have substituted Ti at Ir-site in Y$_2$Ir$_2$O$_7$. The Ti$^{4+}$ (3$d^0$) being a nonmagnetic, one would expect a magnetic dilution and opposite modification of SOC and $U$ parameters accordingly. Therefore, Ti$^{4+}$ sitting at the vertices of Ir tetrahedra will alter the magnetic interaction and electronic charge conduction accordingly. Moreover, Ti$^{4+}$ and Ir$^{4+}$ has comparable ionic radii, thus no major structural modifications would be expected in this series. This being important as structural distortion in the pyrochlore iridates has been shown to have significant influences on the electronic and magnetic properties.\cite{yang}

Our studies show original cubic structural is retained in Y$_2$(Ir$_{1-x}$Ti$_{x}$)$_2$O$_7$ though an evolution of structural parameters have been observed. The parent Y$_2$Ir$_2$O$_7$ shows magnetic irreversibility around 160 K which monotonically decreases with Ti content. Furthermore, both the magnetic moment and the level of frustration decreases with Ti substitution. Y$_2$Ir$_2$O$_7$ shows a nonequilibrium low temperature magnetic state as revealed by magnetic relaxation data, however, relaxation rate increases with Ti. An anomaly in temperature evolution of Raman mode frequency and linewidth is observed around T$_{irr}$ for $x$ = 0.0 and 0.15, however, the magnetic transition at $T_{irr}$ is not accompanied with structural phase transition as confirmed from x-ray diffraction data. The insulating behavior in Y$_2$Ir$_2$O$_7$ is retained with Ti substitution. The charge conduction follows the power-law behavior for each sample in present series. However, we find that the temperature range for validity of power-law behavior changes with compositions.

\begin{figure}[t!]
	\centering
		\includegraphics[width=8cm]{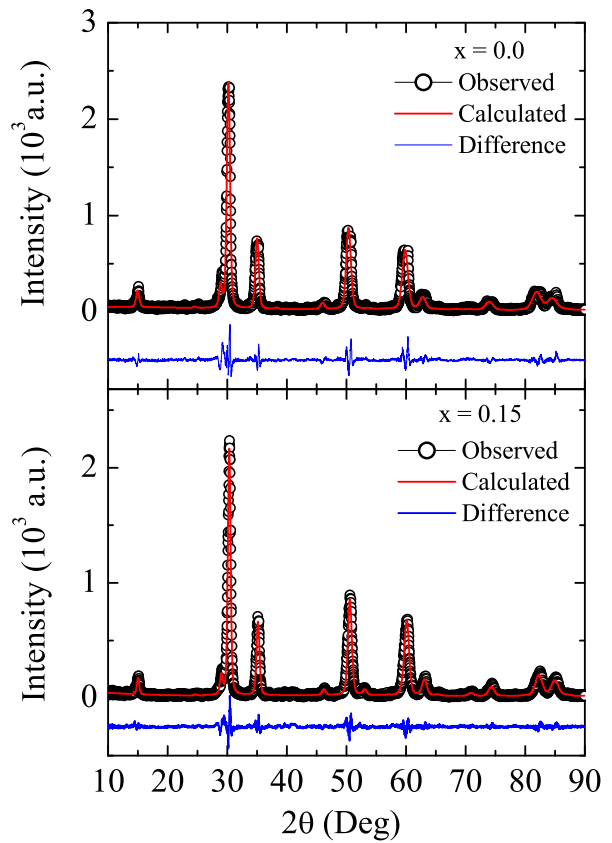}
	\caption{Room temperature XRD pattern along with Rietveld analysis are shown for Y$_2$(Ir$_{1-x}$Ti$_{x}$)$_2$O$_7$ with (a) $x$ = 0.0 and (b) $x$ = 0.15 composition.}
	\label{fig:Fig1}
\end{figure}

\section {Experimental Details}
Polycrystalline samples of Y$_2$(Ir$_{1-x}$Ti$_{x}$)$_2$O$_7$ series with $x$ = 0.0, 0.02, 0.05, 0.10 and 0.15 have been prepared by solid state reaction following the method described in earlier reports. \cite{harish,kumar} The structure and phase purity of these materials have been characterized by powder x-ray diffraction (XRD) using a Rigaku made diffractometer with CuK$_\alpha$ radiation. XRD data are collected in the range of 2$\theta$ = 10 - 90$^o$ at a step of $\Delta 2\theta$ = 0.02$^o$ and with a scan rate of 2$^o$/min. Temperature dependent XRD data have been collected using a PANalytical X'Pert powder diffractometer at different temperatures in the range of 20 - 300 K. The low temperature is achieved using a helium close cycle refregerator (CCR) based cold head where proper temperature stabilization is ensured by waiting sufficiently before collecting the data. The collected XRD data have been analyzed using Rietveld refinement program (FULLPROF) by Rodriguez \textit{et al}.\cite{Rodriguez} DC Magnetization ($M$) measurements have been carried out with a vibrating sample magnetometer (PPMS, Quantum Design). Temperature dependent Raman spectra have been collected using Diode based laser ($\lambda$ = 473 nm) coupled with a Labram-HR800 micro-Raman spectrometer. Electrical transport properties have been measured using a home-built insert fitted with Oxford superconducting magnet. 

\section{Results and Discussions}
\subsection {Structural analysis}
Figure 1(a) and (b) show room temperature XRD data along with Rietveld analysis for the parent with $x$ = 0 and for highest doped material with $x$ = 0.15, respectively. The XRD pattern for the parent material is already reported in our earlier studies,\cite{harish,kumar,hkumar} and this closely match with other studies.\cite{disseler,shapiro,zhu} The Rietveld refinement of the XRD data shows Y$_2$Ir$_2$O$_7$ crystallizes in \textit{Fd$\bar{3}$m} cubic symmetry with lattice constant $a$ = 10.2445 \AA. The analysis of XRD data for present series suggest that the same crystallographic structure and symmetry is retained with Ti substitution which is also expected from the matching ionic radii of Ir$^{4+}$ (0.625 \AA) and Ti$^{4+}$ (0.605 \AA). The ratio $R_{wp}$/$R_{exp}$, which is known as goodness of fit (GOF), is found to be around 1.96, 1.95, 1.92, 1.85, and 1.46 for x = 0.0, 0.02, 0.05, 0.1 and 0.15, respectively, which is indicative of good fitting. The Rietveld analysis indicates an evolution of structural parameters with $x$ as shown in Fig. 2. The unit cell lattice parameter $a$ decreases with Ti concentration (Fig. 2a). This change in $a$ is not substantial as we calculate the change $\Delta a$ is about -0.34 \% over the series. The decrease in the value of lattice parameter with Ti concentration can be explained by the reduced ionic size of Ti$^{4+}$ compared to Ir$^{4+}$.

\begin{figure}
	\centering
		\includegraphics[width=6cm]{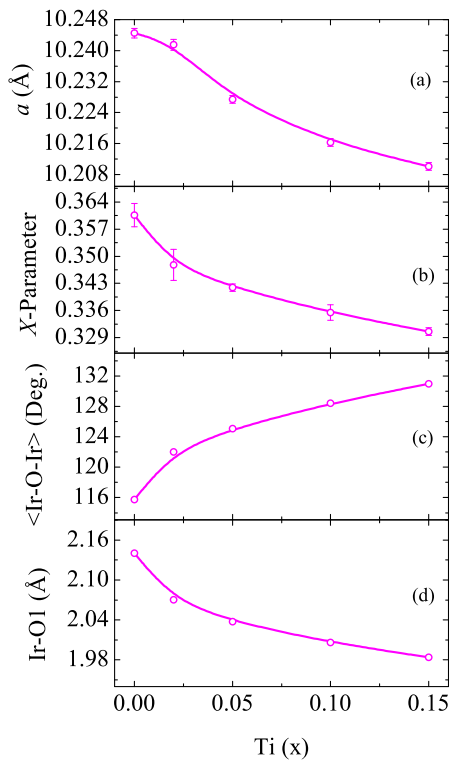}
	\caption{(a) Lattice constant $a$, (b) $X$- parameter, (c) Ir-O-Ir bond angle and (d) Ir-O bond length are shown as a function of Ti substitution for Y$_2$(Ir$_{1-x}$Ti$_{x}$)$_2$O$_7$ series. Lines are guide to eyes.}
	\label{fig:Fig2}
\end{figure}

The variation in structural parameters play an important role in determining the electronic and magnetic properties of a material, in general. There are two adjustable parameters in pyrochlore structure, lattice constant ($a$) and the $x$ coordinate of oxygen site ($X$). Apart from the cubic lattice parameter $a$, the structural organization of IrO$_6$ octahedra is also important. In present pyrochlore structure, the IrO$_6$ octrahedra are connected through oxygen atom i.e., corner oxygen atom is shared between adjacent octahedra. Moreover, in individual octahedra, the six Ir-O bonds have equal length. However, the position of oxygen provides a measure of the distortion of the IrO$_6$ octahedra. For instance, with $X_{ideal}$ = 0.3125 implies octahedra are perfect and undistorted where Ir ions are under the perfect cubic field. Deviation of $X$ from this ideal value generates a distortion to octahedra which lowers the cubic symmetry and induces trigonal crystal field.\cite{JP,Hozo} The $X$ value for this series is shown in Fig. 2b. For Y$_2$Ir$_2$O$_7$, the $X$ value has been found to be 0.36 which implies IrO$_6$ octahedra are distorted and compressed. With substitution of Ti, $X$ value decreases with increasing Ti substitution. This decrease of $X$ value suggests that system is heading toward perfect octahedra, hence trigonal crystal field is reduced. The Ir-O-Ir bond angle and the Ir-O bond length have also significant role on orbital overlapping and charge transfer mechanism, thus those contribute to the electronic properties enormously. The straightening of Ir-O-Ir bond with angle value $\sim$ 180$^o$ realizes a situation where two corner sharing IrO$_6$ octahedra have two IrO$_4$ plaquettes on the same plane and favors maximum electrons hopping between two iridium atoms via oxygen atom. The bending of the Ir-O-Ir bond consequently narrows down the bandwidth and leads to smaller hopping of the electrons along Ir-O-Ir channel. The schematic of Ir-O bond angle and length along with IrO$_6$ octahedra is shown in Fig. 3. Similarly, a shorter Ir-O distance can enhance the overlapping between Ir(5\textit{d}) and O(2\textit{p}) orbitals, leading to larger the hopping of the Ir electrons. An evolution of Ir-O-Ir bond angle and Ir-O bond length for present series is shown in Fig. 2c and d, respectively. With the substitution of Ti, bond-length and bong angle decreases and increases, respectively where both structural modifications are in favor of maximizing the orbital overlapping, hence the electron transport mechanism.

\begin{figure}
	\centering
		\includegraphics[width=7cm]{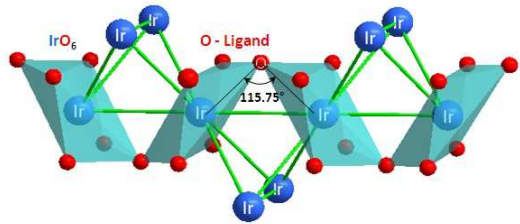}
	\caption{ Structural arrangements of Ir (larger Blue spheres) and O atoms (smaller red spheres) with Ir-O-Ir bond angle within pyrochlore iridate Y$_2$Ir$_{2}$O$_7$. An illustration of the cornered shared Ir$_4$ tetrahedra and oxygen (O) ligands forming IrO$_6$ octrahedra is shown around Ir atoms.}
	\label{fig:Fig3}
\end{figure}

\subsection{Magnetization study}
Temperature dependent magnetization data $M(T)$ measured in 1000 Oe field under the zero field cooled (ZFC) and field cooled (FC) protocol are shown in Fig. 4a for the series Y$_2$(Ir$_{1-x}$Ti$_{x}$)$_2$O$_7$. It is seen in Figure, the ZFC and FC branches of magnetization for parent compound exhibits a clear magnetic irreversibility around $T_{irr}$ $\sim$ 160 K, below which a large bifurcation between magnetization branches is observed.\cite{harish} It can be noted that $M_{ZFC}$ does not show any cusp or peak around the $T_{irr}$. Given that low temperature magnetic state in Y$_2$Ir$_2$O$_7$ is debated showing possible presence of long-range magnetic ordering or nonequilibrium ground state,\cite{disseler,shapiro,harish,kumar} the dilution of magnetic lattice with substitution of nonmagnetic Ti$^{4+}$ will be quite interesting. As evident in Fig. 4a, the notable observations are, bifurcation temperature $T_{irr}$ shifts toward low temperature, difference between $M_{FC}$ and $M_{ZFC}$ decreases and the low temperature magnetic moment decreases. The inset of Fig. 4a shows $M_{ZFC}$ moment at 5 K initially rapidly decreases with Ti doping, however, for higher value of $x$ ($> 5\%$) the decrease is not substantial. Similarly, the variation of $T_{irr}$ with $x$ is shown in the Fig. 4b which shows $T_{irr}$ decreases only by $\sim$ 8 K with 15\% of Ti substitution. This is rather surprising with present nonmagnetic Ti$^{4+}$ as previous study with magnetic substitution in Y$_2$(Ir$_{1-x}$Ru$_{x}$)$_2$O$_7$ shows substantial decrease of $T_{irr}$ ($\sim$ 40 K) with similar doping concentration.\cite{kumar} 

\begin{figure}[t!]
	\centering
		\includegraphics[width=8.5cm]{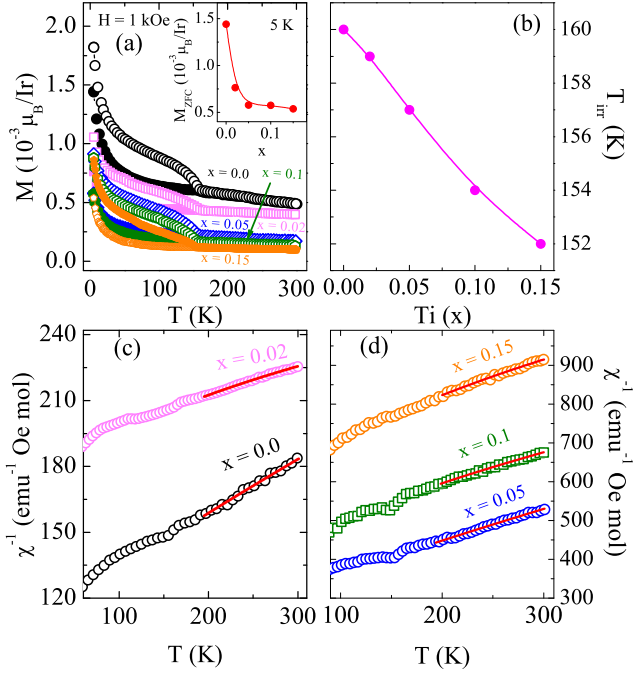}
	\caption{(a) Temperature dependent magnetization data measured in 1 kOe field under ZFC and FC protocol are shown for Y$_2$(Ir$_{1-x}$Ti$_{x}$)$_2$O$_7$ series. Inset shows the variation of $M_{ZFC}$ moment with $x$ at 5 K. (b) shows variation of transition temperature $T_{irr}$ with Ti substitution for the same series. Temperature dependent inverse susceptibility ($\chi^{-1}$ = $(M/H)^{-1}$) are shown with (c) $x$ = 0.0, 0.02 and (d) $x$ = 0.05, 0.10 and 0.15 for Y$_2$(Ir$_{1-x}$Ti$_{x}$)$_2$O$_7$ series. The solid lines are due to fitting using Eq. 1.}
	\label{fig:Fig4}
\end{figure}

\begin{figure}[t!]
	\centering
		\includegraphics[width=6cm]{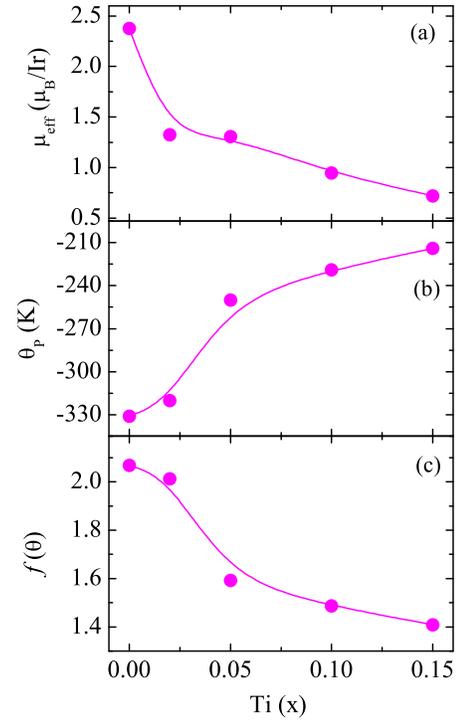}
	\caption{(a) Effective paramagnetic moment (b) Curie-Weiss temperature (c) Frustration parameter are shown with doping concentration $x$ in Y$_2$(Ir$_{1-x}$Ti$_{x}$)$_2$O$_7$ series. These parameters are obtained from fitting of magnetization data with Eq. 1.}
	\label{fig:Fig5}
\end{figure}

To understand the magnetic state in further detail, we have plotted inverse susceptibility ($\chi^{-1} = (M/H)^{-1}$) as a function of temperature for this series in Fig. 4c ($x$ = 0.0, 0.02) and 4d ($x$ = 0.05, 0.1, 0.15). As evident, the $\chi^{-1}(T)$ data almost show linear behavior above $T_{irr}$. The $\chi^{-1}(T)$ data have been fitted in high temperature range (200 - 300 K) with modified Curie-Weiss law;

\begin{eqnarray}
	\chi = \chi_0 + \frac{C}{T - \theta_P}
\end{eqnarray}

where $\chi_0$, $C$, and $\theta_P$ are the temperature independent magnetic susceptibility, the Curie constant, and Curie-Weiss temperature, respectively. For $x$ = 0.0 parent material, the obtained fitted parameters are $\chi_0$ = 9.7 $\times$ 10$^{-4}$ emu/mole, $C$ = 2.83 emu K/mole and $\theta_P$ = -331 K.\cite{kumar} Using the fitted parameter $C$, we have calculated the effective paramagnetic moment $\mu_{eff}$ = 2.37 $\mu_B$/Ir for $x$ = 0.0 which appears higher than the expected value ($g\sqrt{S(S+1)} \mu_B$) 1.73 $\mu_B$/Ir in case of spin-only value ($g$ = 2 and $S$ = 1/2). Fig. 5(a) shows the composition dependent $\mu_{eff}$ for Y$_2$(Ir$_{1-x}$Ti$_{x}$)$_2$O$_7$ series. As evident in figure, the value of $\mu_{eff}$ decreases with Ti substitution. Ti is assumed to adopt Ti$^{4+}$ (3$d^0$, $S$ = 0) which implies a decrease of $\mu_{eff}$ in present series. The composition dependent $\theta_P$ is plotted in Fig. 5b shows its magnitude decreases with $x$. The sign as well as magnitude of $\theta_P$ suggest an antiferromagnetic type interaction type magnetic interaction with reasonable strength. The nature of evolution of $\theta_P$ underlines the fact that nature of magnetic interaction remains to be antiferromagnetic type interaction but its strength is weakened with introduction of Ti. We have further calculated the relevant frustration parameter $f$ which is defined as the ratio $\left|\theta_P\right|$/$T_{irr}$. These pyrochlore structures are inherently frustrated due to their geometrical configuration and often leads to different magnetic ground states such as, spin-glass\cite{gingras,yoshii}, spin-liquid \cite{gardner1,nakatsuji}, spin-ice \cite{bramwell,fukazawa1}, etc. For example, corresponding $f$ values for highly frustrated pyrochlore oxides such as, Y$_2$Ru$_2$O$_7$ and Y$_2$Mo$_2$O$_7$ are about 16 and 8.8, respectively which exhibit spin-glass behavior.\cite{kmiec,Gardner}. Fig. 5 (c) shows the $f$ is reasonable in parent compound though decreases with Ti. It is interesting to observe in Fig. 5 that all the parameters ($\mu_{eff}$, $\theta_P$ and $f$) exhibit rapid changes with initial doping concentration of Ti till $x$ = 0.05 and then a slow evolution at higher Ti doping which is also consistent with $M_{ZFC}$ value at 5 K (inset of Fig. 4a). 

\begin{figure}[t!]
	\centering
		\includegraphics[width=6.5cm]{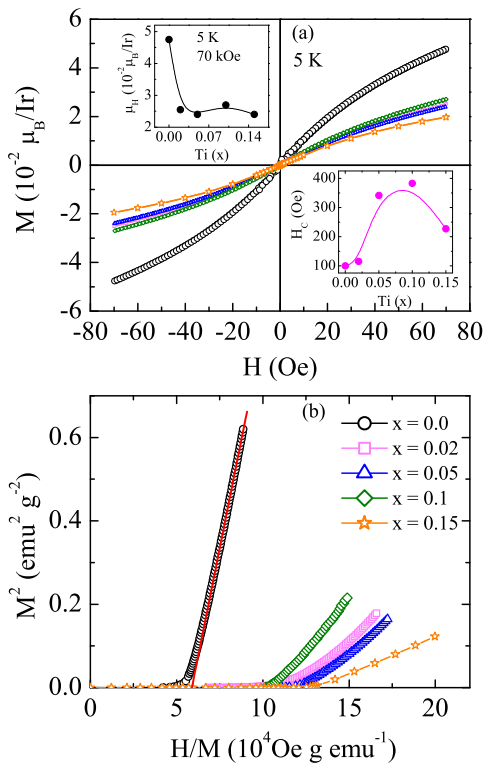}
	\caption{(a) Magnetic field dependent magnetization are shown for Y$_2$(Ir$_{1-x}$Ti$_{x}$)$_2$O$_7$ at 5 K. Upper and lower inset show composition dependent moment $\mu_H$ at field 70 kOe and coercive field $H_c$, respectively. (b) The M(H) data are plotted in the form of Arrott plot ($M^2$ vs $H/M$) for the same materials.}
	\label{fig:Fig6}
\end{figure}

The magnetic field dependent of magnetization data collected at 5 K in the field range of $\pm$ 70 kOe have been shown in Fig. 6a. As evident in figure, the $M(H)$ data for the $x$ = 0.0 parent material does not show linear behavior and the magnetization increases continuously without showing any saturation till 70 kOe. With Ti substitution, the moment $\mu_H$ at high field decreases though not linearly as expected from site dilution with nonmagnetic element. The upper inset of Fig. 6a shows variation of $\mu_H$ with $x$. Further, a close observation of $M(H)$ data shows a small hysteresis with coercive field $H_c$ $\sim$ 100 Oe at 5 K for Y$_2$Ir$_2$O$_7$. The $H_c$ initially increases with $x$ and then decreases at higher doping (lower inset in Fig. 6a) which is indicative of increasing FM behavior. To understand the nature of magnetic state in further detail, we have plotted the $M(H)$ data in form of Arrott plot ($M^2$ vs $H/M$).\cite{arrott} Fig. 6b shows the Arrott plot of $M(H)$ data as shown in Fig. 6a. The detail significance of Arrott plot is discussed elsewhere.\cite{harish,kumar} As evident in Fig. 6b, the intercept of straight line fitting turns out to be negative for all the materials, though we find the negative value decreases with Ti doping. This indicates low temperature magnetic state in the Y$_2$Ir$_2$O$_7$ is not FM type, and Ti substitution reduces the non-FM character in this series which is also supported by the fact that $H_c$ increases in this series.

Above discussion of magnetic behavior in Y$_2$(Ir$_{1-x}$Ti$_{x}$)$_2$O$_7$ series imply that both the magnetic moment as well as level of frustration decreases with Ti which is quite interesting (Figs. 4a, 5a and Fig. 5c). The substitution of nonmagnetic Ti$^{4+}$ (3$d^0$) for Ir$^{4+}$ (5$d^5$) will unlikely introduce its own magnetic interaction and it will act for site dilution. Due to high crystal field splitting in 5$d$ materials, all the five $d$ electrons will occupy the low energy $t_{2g}$ level. It is commonly believed that the strong SOC effect further splits the $t_{2g}$ level into $J_{eff}$ = 1/2 doublet and $J_{eff}$ = 3/2 quartet where the latter has low energy.\cite{kim} This gives a situation of partially filled $J_{eff}$ = 1/2 state where the Fermi level E$_F$ resides. In picture of strong SOC, therefore, Ir$^{4+}$ gives moment ($g_J$$J_{eff}$$\mu_B$) value 0.33 $\mu_B/Ir$. This further argues that in presence of on-site Coulomb repulsion $U$, a Mott-like gap is driven in the $J_{eff}$ = 1/2 state which makes these materials $J_{eff}$ insulator. The suppression of the magnetic moment in Y$_2$(Ir$_{1-x}$Ti$_{x}$)$_2$O$_7$ can be explained as Ti will simply dilute the observed magnetic moment. The similar weakening of magnetic ordering and insulating state has previously been observed in Y$_2$Ir$_2$O$_7$ with magnetic Ru$^{4+}$ substitution \cite{kumar}, however, effect of Ru appears stronger than Ti. The Ti being a 3$d$ element has comparatively higher electron correlation $U$ effect than Ru. Whether this plays a role for small decrease of $T_{irr}$ in Y$_2$(Ir$_{1-x}$Ti$_{x}$)$_2$O$_7$ need to be investigated throughly. Nonetheless, Ru doping in different iridate materials has shown a common trend of suppressing magnetic state.\cite{kumar,cava,yuan,calder} Note, that Ti substitution in other Ir based materials has shown similar rather interesting results. For example, in Sr$_{2}$IrO$_{4}$ the moment is seen to decrease with progressive Ti content.\cite{Gatimu,ge} In case of Na$_{2}$Ir$_{1-x}$Ti$_{x}$O$_{3}$, the similar weakening of magnetic ordering is observed, though the system evolved to a glassy state, indicating a presence of strong frustration.\cite{Manni} Apart from strong SOC picture, recent theoretical studies have further considered the possible influence of non-cubic crystal field such as, trigonal crystal field on the magnetic and electronic properties in iridium oxides.\cite{yang,liu} Whether such effects are responsible for present evolution of magnetic behavior needs to be investigated.

\begin{figure}
	\centering
		\includegraphics[width=8.5cm]{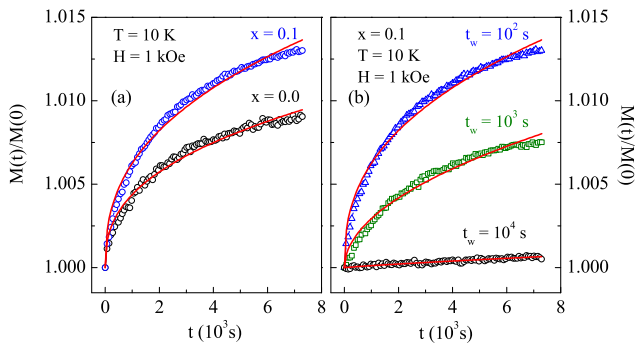}
	\caption{(Color online) a) The normalized magnetic moment $M(t)/M(0)$ as a function of time have been shown for $x$ = 0.0 and 0.1 with $t_w$ = 10$^2$ s. (b) The similar normalized moment have been shown for different wait time $t_w$ for doped Y$_2$(Ir$_{1.9}$Ti$_{0.1}$)$_2$O$_7$. The solid lines show representative fitting of data using Eq. 2.}
	\label{fig:Fig7}
\end{figure}
\begin{table}[b!]
\caption{\label{label} The characteristic relaxation time $\tau$ and the exponent $\beta$ as obtained from fitting of magnetic relaxation data (Fig. 7) with Eq. 2 are shown for different waiting time $t_w$ for $x$ = 0.0 and 0.1 sample of Y$_2$(Ir$_{1-x}$Ti$_{x}$)$_2$O$_7$ series.}
\begin{ruledtabular}
\begin{tabular}{cccc}
\hline
$t_w (s)$ &$x$ &$\tau$ (s) &$\beta$\\      
\hline
10$^2$  &0.0 &5.6 $\times$ 10$^9$ &0.33(1)\\
   &0.1 &4.4 $\times$ 10$^8$ &0.39(1)\\
\hline
10$^3$  &0.0 &1.78 $\times$ 10$^9$ &0.41(2)\\
  &0.1 &1.07 $\times$ 10$^8$ &0.50(2)\\
\hline
10$^4$  &0.0 &0.10 $\times$ 10$^9$ &0.82(8)\\
  &0.1 &0.41 $\times$ 10$^8$ &0.85(2)\\
\hline
\end{tabular}
\end{ruledtabular}

\end{table}
Pyrochlore structures are inherently frustrated due to its geometrical arrangement of magnetic atoms on vertices of tetrahedra. This frustration generates various metamagnetic states including glassy behavior. While for classical magnetic systems the magnetization normally depends only on temperature and magnetic field. However, in metamagnetic states the moment shows an unusual evolution with time ($t$) even sample is held at same temperature and field. Additionally, time evolution of magnetic moment shows a dependence on history how magnetic field is applied, a phenomenon called as aging behavior. Recently, we have reported nonequilibrium magnetic state in Y$_2$Ir$_2$O$_7$ at low temperature showing reasonable magnetic relaxation and aging behavior.\cite{harish} To understand the evolution of relaxation and aging behavior with Ti substitution, we have measured $M(t)$ following a protocol where the material is cooled in zero field from room temperature to 10 K. After stabilization of temperature, a magnetic field of 1000 Oe is applied after a wait time $t_w$ and then magnetic moment is measured as a function of time for about 7200 s. Fig. 7a shows normalized relaxation data at $t$ = 0 with time for two representative samples i.e., $x$ = 0.0 and 0.1 of present series. We find that the magnetization ($M(t)/M(0)$) increases continuously without any saturation for both the materials, however, the relaxation rate is higher in doped sample. This magnetic relaxation in these compounds suggests that the low temperature magnetic state is basically a nonequilibrium state as it tries to achieve low energy state(s) with time crossing the barriers which separate the energy states.

We have further measured magnetic relaxation measurement with different $t_w$, as presented in Fig. 7b for $x$ = 0.1 sample. For the aging measurement, the sample is similarly cooled in zero field and a magnetic field of 1000 Oe is applied after certain $t_w$. During wait time the system relaxes toward the equilibrium state. The $M(t)$ data have been shown for $t_w$ = 10$^2$, 10$^3$ and 10$^4$ s. The normalized relaxation data show relaxation rate diminishes significantly with increasing $t_w$ which implies that the system waits more time at 10 K before applying the field shows less relaxation, thus the magnetic relaxation depends on history how the magnetic field is applied even at the same temperature and with same magnetic field. This indicates that during the wait time, system ages and tries to achieve the low energy state which is evident from low relaxation rate in Fig. 7b.

For quantitative analysis of relaxation data, we find that $M(t)$ data can be best explained with stretched exponential function.\cite{chamberlin}

\begin{eqnarray}
	M(t) = M(0)\exp \left(\frac{t}{\tau}\right)^\beta
\end{eqnarray}
 
where $\beta$ is the stretching exponent with 0 $< \beta \leq$ 1, $M(0)$ is the magnetization at $t$ = 0 and $\tau$ is the characteristic relaxation time. The stretched exponential form of magnetic relaxation explains well the system which has multiple energy barrier. The solid lines in Figs. 7a and 7b show the fitting of our data using Eq. 2 for different samples and different $t_w$. The parameters $\tau$ and $\beta$ obtained from the fitting are shown in Table 1 for $x$ = 0.0 and 0.1 compound. For $t_w$ = 10$^2$ s, $\tau$ decreases by an order while $\beta$ increases in doped sample (Table 1). This decrease of $\tau$ and simultaneous increase of exponent $\beta$ with in doped sample is in agreement with higher relaxation behavior as evident in Fig. 7a. Table 1 further shows this pattern is maintained for all $t_w$. To understand the effect of $t_w$, Table 1 shows with increasing $t_w$ the $\tau$ decreases where $\beta$ increases for both the samples which are consistent with slow relaxation with increasing $t_w$ (Fig. 7b). This clearly indicates doped Ti somehow helps the spin system to attain the low-energy equilibrium state at faster rate which is also supported by reduced frustration parameter in doped samples (Fig. 5c).

\begin{figure}
	\centering
		\includegraphics[width=7cm]{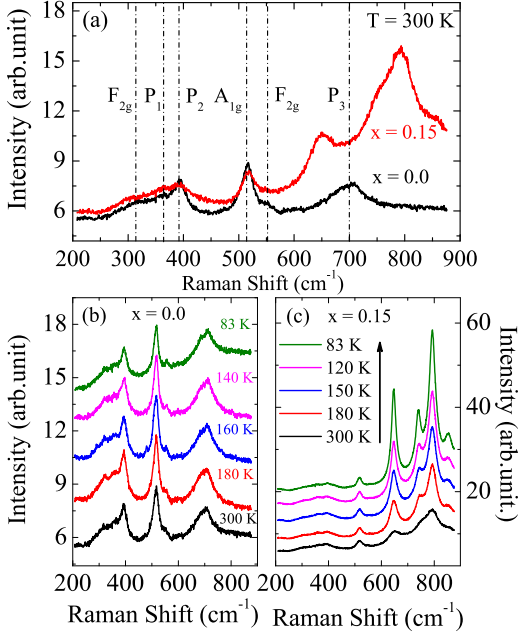}
	\caption{(Color online) (a) Raman spectra measured at room temperature are shown for Y$_2$(Ir$_{1-x}$Ti$_{x}$)$_2$O$_7$ with $x$ = 0.0 and 0.15. Vertical lines correspond to different Raman modes. (b) and (c) show Raman spectra at selected temperatures for Y$_2$(Ir$_{1-x}$Ti$_{x}$)$_2$O$_7$ with $x$ = 0.0 and 0.15, respectively. The data at different temperatures are shifted vertically for clarity.}
	\label{fig:Fig8}
\end{figure}

\subsection{Temperature dependent Raman mode study}
The structural information in present series has further been obtained using Raman spectrum, which is generally sensitive to oxygen-cation vibrations and the identity of pyrochlore structure. The Y$_2$Ir$_2$O$_7$ belongs to the cubic-\textit{Fd$\bar{3}$m} structural symmetry which has six Raman active modes that are expected according to group factor, $\Gamma^R$ = A$_{1g}$ + E$_g$ + 4F$_{2g}$. \cite{han,hasegawa,Jurel} These Raman modes involve only the motion of oxygen atoms. Fig. 8a shows Raman spectra collected for two end members of Y$_2$(Ir$_{1-x}$Ti$_{x}$)$_2$O$_7$ series with $x$ = 0.0 and x = 0.15 at room temperature. For Y$_2$Ir$_2$O$_7$, six Raman active modes are observed in Fig. 8a which are consistent with the previous Raman data for pyrochlore iridates.\cite{han,hasegawa,Jurel} We observe F$_{2g}$, P$_1$, P$_2$, A$_{1g}$, F$_{2g}$ and P$_3$ modes appear at 314, 361, 393, 516, 552, and 700 cm$^{-1}$. The A$_{1g}$ mode at 516 cm$^{-1}$ has an important role in pyrochlore structure because it is related to the trigonal distortion of IrO$_6$ octrahedra as well as with the changes of Ir-O-Ir bond angle.\cite{Taniguchi} As discussed, trigonal distortion of IrO$_6$ octrahedra mainly arises due to the change in value of $X$ parameter of oxygen atom (Fig. 2b). Therefore, A$_{1g}$ mode is directly involved in the modulation of $X$ parameter of the oxygen atom residing at 48$f$ site.\cite{Sanjuan, tracy} Similarly, E$_g$ mode is assigned to IrO$_6$ bending vibrations and F$_{2g}$ relates to the mixture of Y-O and Ir-O bond stretching vibrations.\cite{sreena} In previous study, the peak P$_2$ at 392 cm$^{-1}$ is assigned as E$_g$ or F$_{2g}$.\cite{hasegawa} Fig. 8b shows the Raman spectrum for Y$_2$Ir$_2$O$_7$ at temperatures 300, 180, 160, 140 and 83 K where the temperatures represent the room temperature PM state (300 K), above (180 K) and below (140 K) the $T_{irr}$ and low temperature state (83 K). It is clear from the figure that Raman spectra does not show any change with temperature in terms of appearing of new mode(s) or mode shifting. This implies that there is no change in structural symmetry with temperature in agreement with the temperature dependent of XRD and neutron scattering measurements.\cite{harish,shapiro} On other hand, Raman spectra for $x$ = 0.15 material at room temperature (Fig. 8a) shows two new additional peaks at 648, and 792 cm$^{-1}$ and two humps at 748, and 855 cm$^{-1}$, however, one original peak at 700 cm$^{-1}$ is not evident in doped $x$ = 0.15 system. These new peaks are commonly found in Ti/Ir based pyrochlores as an intense and broad peak in the frequency range 640 to 900 cm$^{-1}$,\cite{Sanjuan,tracy} and are often assigned to the second order Raman scattering of pyrochlore oxides.\cite{Saha,hasegawa,Jurel}

\begin{figure}
	\centering
		\includegraphics[width=8cm]{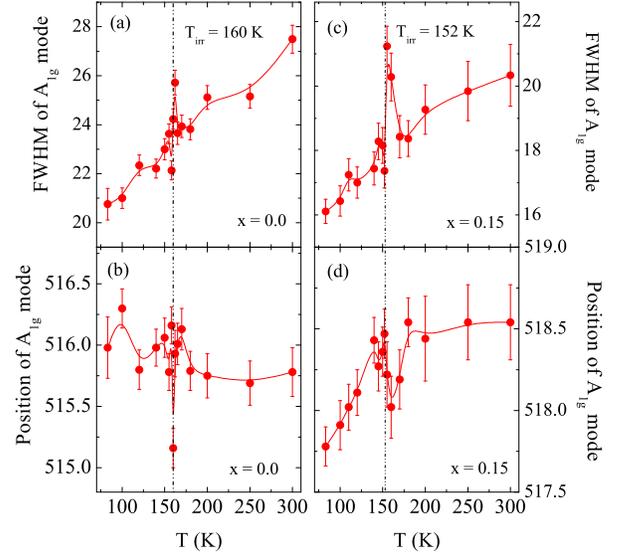}
	\caption{(Color online) (a) and (c) show the full width at half maximum (FWHM), and (b) and (d) show the frequency of A$_{1g}$ phonon mode as a function of temperature for $x$ = 0.0 and 0.15, respectively. Vertical lines represent the respective $T_{irr}$.} 
	\label{fig:Fig9}
\end{figure}

Temperature dependent Raman spectra for $x$ = 0.15 sample are shown in Fig. 8c. With decreasing temperature, while there is no change in spectra in low frequency modes but the four new peaks/humps at 648, 748, 792 and 855 cm$^{-1}$ become more intense and prominent at low temperature. Given that role of A$_{1g}$ mode in pyrochlore structure is important, we have analyzed this mode of Raman data in detail. The evolution of line width as represented by full width at half maximum (FWHM) and the position of A$_{1g}$ mode are shown with temperature in Fig. 9 for $x$ = 0.0 and 0.15. For $x$ = 0.0, Fig. 9a depicts the FWHM of A$_{1g}$ peak decreases with the decreasing temperature where the change in FHWM is $\sim$ - 24\% over this temperature range. It is notable that a distinct peak in FWHM around $T_{irr}$ is observed. The changes of A$_{1g}$ peak position with temperature is presented in Fig. 9b which shows some fluctuations across $T_{irr}$. Given that A$_{1g}$ mode is directly related to the oxygen $X$ parameter, therefore, the observed fluctuations in Fig. 9b is quite consistent with the $X$ parameter fluctuation of this materials as reported previously.\cite{harish} The FWHM and position of A$_{1g}$ mode for the Ti-doped material ($x$ = 0.15) are shown in Figs. 9c and 9d, respectively. The results are quite similar with the parent material in terms of FWHM and frequency fluctuation of A$_{1g}$ peak around respective $T_{irr}$. This anomaly in Raman data may possibly arise due to sudden rotation/distortion of IrO$_6$ octahedra across $T_{irr}$. Compared to the parent system, the phonon parameters for the Ti-doped material ($x$ = 0.15) are shifted to low temperature by 8 K, consistent with a suppression of magnetic ordering with Ti doping ($x$ = 0.15) in parent material. Similar evolution of phonon parameters with magnetic ordering has been reported in its Ru doped analogues Y$_2$(Ir$_{1-x}$Ru$_{x}$)$_2$O$_7$ and doped iridate Sr$_2$Ir$_{1-x}$Ru$_x$O$_4$.\cite{Jurel,glama,cetin}

\begin{figure}[t!]
	\centering
		\includegraphics[width=8cm]{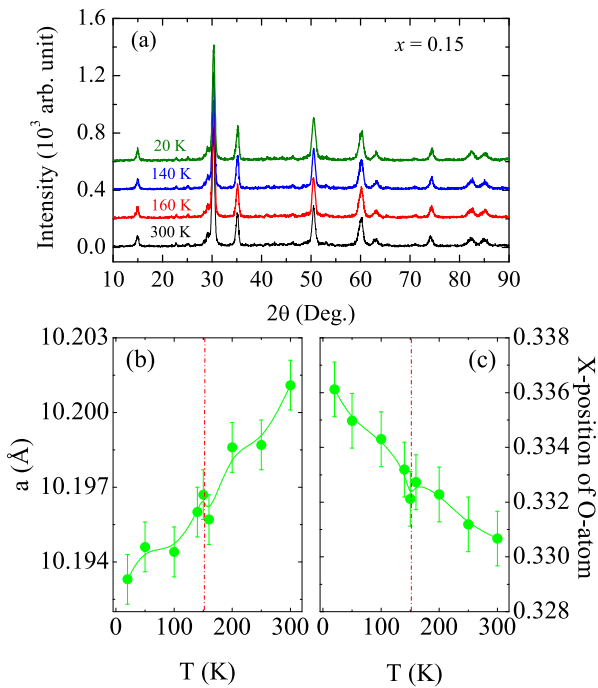}
	\caption{(Color online) (a) Temperature dependent XRD data are shown for Y$_2$(Ir$_{1.85}$Ti$_{0.15}$)$_2$O$_7$ at selective temperatures 300, 160, 140, and 20 K. Temperature variation of lattice parameters (a) $a$ and (b) $X$ - parameter of O-atom are shown for Y$_2$(Ir$_{1.85}$Ti$_{0.15}$)$_2$O$_7$. Vertical dotted lines signify the $T_{irr}$ of this material.} 
	\label{fig:Fig10}
\end{figure}

\subsection{Temperature dependent structural investigation}
Previous studies imply a strong correlation between the magnetic behavior and the Raman data, therefore it is necessary to understand the evolution of structural parameters with temperature particularly across the $T_{irr}$. Previous XRD study demonstrate that there is no structural phase transition down to low temperature in Y$_2$Ir$_2$O$_7$ where the $X$ parameter related to oxygen position anomalous fluctuation across the $T_{irr}$.\cite{harish} With this aim, in present study we have done temperature dependent XRD measurements for $x$ = 0.15 material at various temperatures down to 20 K. Fig. 10a shows the representative XRD plots of Y$_2$(Ir$_{1.85}$Ti$_{0.15}$)$_2$O$_7$ at 300, 160, 140 and 20 K where the temperatures represent different magnetic states (Fig. 4). The Fig. 10a shows no major modification in XRD patterns in terms of splitting of the peaks or arising of new peak(s) implying that the magnetic phase transition is not accompanied by any change in the structural symmetry. The XRD patterns are analyzed with Rietveld refinement (not shown) which indicates this material retains its original cubic-\textit{Fd$\bar{3}$m} structural symmetry down to 20 K. Fig. 10b represents the temperature dependent lattice constant $a$ as obtained from the Rietveld analysis. The lattice constant $a$ decreases with decreasing the temperature similar to parent compound.\cite{harish} The $X$ parameter of oxygen atom in IrO$_6$ octahedra though increases but shows a small fluctuation across $T_{irr}$ (Fig. 10c) which is consistent with Raman study where the line width (FWHM) and position of A$_{1g}$ mode show an anomaly around this transition temperature (see Fig. 9).

\subsection{Electrical transport}
The effect of Ti substitution on charge transport in Y$_2$(Ir$_{1-x}$Ti$_{x}$)$_2$O$_7$ is shown in Fig. 11a which demonstrates electrical resistivity $\rho$ as a function of temperature. The Y$_2$Ir$_2$O$_7$ is reported as a strong insulator where the nature of charge transport follows a power-law dependence on temperature.\cite{disseler,harish,kumar,Ramirez} In that respect, Ti substitution which is expected to dilute the magnetic network as well as to tune the both SOC and $U$ will have strong ramification on the electron transport behavior. The $\rho(T)$ for Y$_2$Ir$_2$O$_7$ shows highly insulating behavior where the resistivity increases by couple of orders at low temperature. With Ti doping, all the samples remain insulating while the resistivity shows nonmonotonic behavior (Fig. 11). For instance, resistivity initially decreases substantially up to $x$ = 0.05 and then it increases. The evolution of resistivity with Ti doping at 25 K is shown in inset of Fig. 11a where this behavior arises due to a likely interplay between the Ti induced factors mentioned above. We have previously shown that the nature of charge conduction follows the power-law behavior as 

\begin{eqnarray}
	\rho = \rho_0 T^{-n}
\end{eqnarray}

where $n$ is power exponent. The inset of Fig. 11b shows ln-ln plot and straight line fitting of $\rho(T)$ data following Eq. 3 for the parent compound where the data can be fitted throughout the temperature range with exponent $n$ = 2.98. For the doped samples, however, we can fit the $\rho(T)$ data with Eq. 3 only in low-temperature range as shown in Fig. 11b. The range of straight line fitting for the series extends from low temperature to 52, 21, 37 and 88 K and the obtained $n$ values are 3.04, 3.69, 4.3 and 4.54 for $x$ = 0.02, 0.05, 0.1 and 0.15, respectively. While the exponent $n$ increases continuously, the fitting temperature range follow similar pattern of resistivity value showing minimum range for $x$ = 0.05 sample.

\begin{figure}[t!]
	\centering
		\includegraphics[width=7cm]{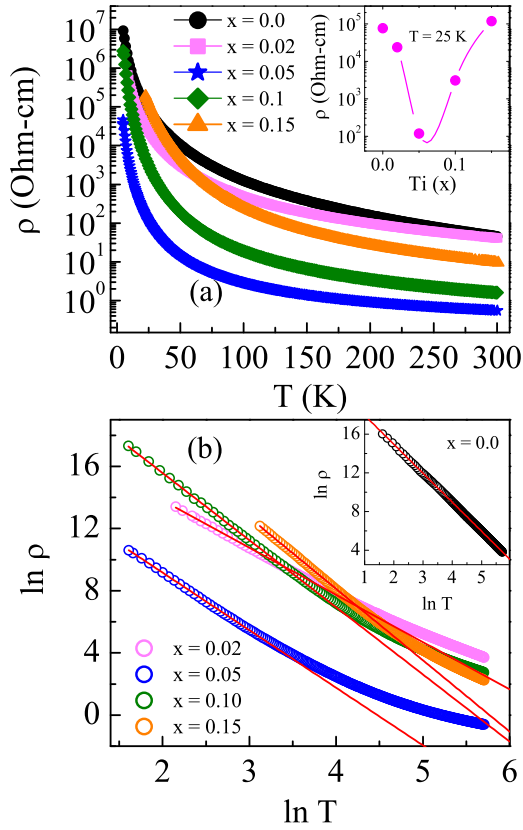}
	\caption{(color online) (a) Temperature dependent resistivity data of Y$_2$(Ir$_{1-x}$Ti$_{x}$)$_2$O$_7$ series has been shown in semi-log plot. Inset shows the variation of resistivity with $x$ at 25 K. (b) The $\rho(T)$ data of the series in (a) are plotted in $\ln-\ln$ plot. The straight lines are due to fitting with Eq. 3. Inset shows the resistivity data for Y$_2$Ir$_2$O$_7$ in $\ln-\ln$ plot.}
	\label{fig:Fig11}
\end{figure}

\begin{figure}[t!]
	\centering
		\includegraphics[width=6.5cm]{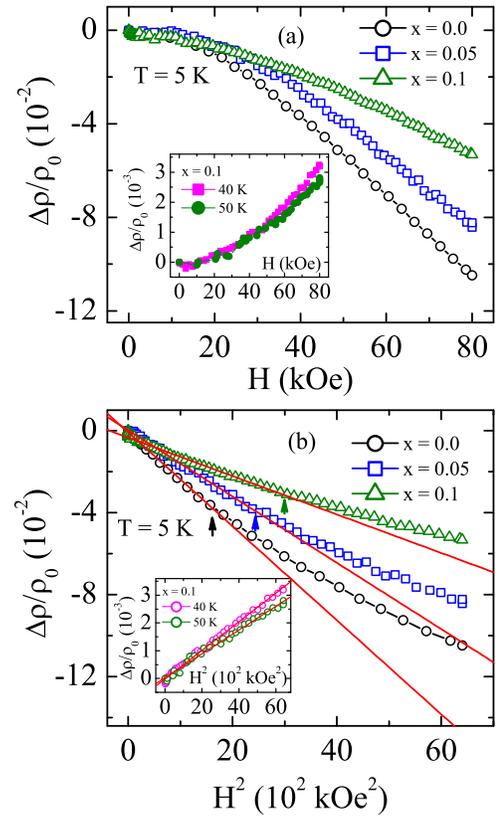}
	\caption{(color online) (a) Magnetoresistance measured at 5 K in a magnetic field up to 80 kOe are shown for Y$_2$(Ir$_{1-x}$Ti$_{x}$)$_2$O$_7$ series. Inset shows MR for $x$ = 0.1 sample at 40 and 50 K showing positive behavior. (b) shows the quadratic field dependance of negative MR at 5 K for Y$_2$(Ir$_{1-x}$Ti$_{x}$)$_2$O$_7$ series. Inset shows similar quadratic field dependence of positive MR for $x$ = 0.1 at 40 and 50 K.}
	\label{fig:Fig12}
\end{figure}

Further, we have studied the electronic transport behavior in presence of magnetic field at low temperature. This adds interest because magnetoresistance (MR) has shown an interesting effect in SOC dominated systems i.e., Bi$_2$Se$_3$,\cite{chen1} Bi$_2$Te$_3$,\cite{he} Au covered Mg film,\cite{berg} even in Ir based Na$_2$IrO$_3$ films,\cite{jender} where the positive MR has mostly been explained with weak antilocalization (WAL) or weak localization (WL) effect. Fig. 12 (a) shows isothermal magnetoresistance measured at 5 K for Y$_2$(Ir$_{1-x}$Ti$_{x}$)$_2$O$_7$ series with $x$ = 0.0, 0.05 and 0.1. Magnetoresistance is understood as a change in the electrical resistance in presence of magnetic field,

\begin{eqnarray}
	\Delta \rho/\rho(0) = \left[\rho(H) - \rho(0)\right]/\rho(0)
\end{eqnarray}

where, $\rho(0)$ and $\rho(H)$ are the resistivity at zero field and presence of magnetic field, respectively. As evident in Fig. 12 (a), all the samples show negative MR at 5 K i.e., the resistivity decreases in presence of magnetic field. The parent Y$_2$Ir$_2$O$_7$ exhibits $\sim$ 11 \% MR at 80 kOe field. The negative MR in SOC dominated systems is mostly theoretically explained with weak localization (WL) effect. With Ti substitution, the general feature of MR remains similar but its magnitude decreases continuously with $x$. For example, at same temperature and field, MR value is found around 8.5 and 5.3 \% for $x$ = 0.05 and 0.1, respectively. Theoretical calculations explain this negative MR in disordered systems through quantum interference effect which causes the localization of charges where the application of magnetic field destroy this interference effect and negative MR is realized. The calculations employing critical path analysis further show a quadratic field dependence of MR.\cite{sivan}

Fig. 12(b) presents MR at 5 K as a function of $H^2$ for selective samples with $x$ = 0.0, 0.05 and 0.1, which shows a quadratic field dependence in low field region. However, field range of quadratic dependence increases with $x$. We observe MR vs. $H^2$ follows linear dependence till field 43, 52 and 55 kOe for $x$ = 0.0, 0.05 and 0.1, respectively. This behavior in Fig. 12(b) implies charge transport in present series follows weak localization mechanism. The MR evolution with both $x$ and magnetic field can be explained with significant weakening of SOC and dilution of charge transport path due to Ti$^{4+}$ ions. This is direct evidence of tuning of SOC with Ti$^{4+}$ in Y$_2$Ir$_2$O$_7$. Following that evolution of MR from negative to positive with temperature in Ir based Sr$_2$IrO$_4$ films,\cite{Miao} we have checked MR with temperature in one selective sample i.e., $x$ = 0.1. Inset of Fig. 12(a) shows MR exhibits positive value in high temperature at 40 and 50 K. This crossover from negative to positive MR is interesting and may be due to an interplay between magnetic field and magnetic moment where the increasing value of the later induces negative MR at low temperature. We find the magnitude of positive MR is almost one order lower than its negative component, as observed for Sr$_2$IrO$_4$ films.\cite{Miao} The positive MR does not exhibit any cusp or peak close to zero field which excludes the possibility of weak-antilocalization effect as has been observed in other Ir based oxides such as, Na$_2$IrO$_3$ films.\cite{jender} We, however, find that positive MR follows quadratic field dependence till the highest measuring field of 80 kOe which is in contrast with negative MR that shows $H^2$ dependence in only low field regime (Fig. 12 (b)).

The insulating states in Ir-based materials are of general interest which have been investigated using different substitutions. In case of doped layered Sr$_2$Ir$_{1-x}$Ti$_x$O$_4$, the resistivity is increases considerably where the another end member i.e., Sr$_2$TiO$_4$ is highly insulating.\cite{Gatimu,ge} The changes of resistivity in present Y$_2$(Ir$_{1-x}$Ti$_{x}$)$_2$O$_7$ is rather interesting. Along with the dilution of Ir-O-Ir connectivity and tuning the SOC and $U$, the structural modification due to Ti also need to be considered for resistivity analysis. As seen in Fig. 2, the Ir-O bond length and angle change continuously with Ti doping which promotes an orbital overlapping of Ir-$d$ and O-$p$ orbitals. This likely to facilitate the electronic transfer, hence decreasing the resistivity of material in initial doped samples. However, other competing factors such as, site dilution and possible role of opposing tuning of SOC and $U$ realized through Ti substitution may contribute to increase of resistivity in higher doped samples.

From the above discussion it is clear that Ti substitution has interesting effects on structural, magnetic and electronic transport behavior in Y$_2$(Ir$_{1-x}$Ti$_{x}$)$_2$O$_7$ series. While the magnetic moment in present series decreases but the ordering temperature $T_{irr}$ shows an insignificant change with $x$ (Fig. 4). It can be mentioned that Ti$^{4+}$ doping shows a weaker effect on $T_{irr}$ compared to magnetically active Ru$^{4+}$ doping in Y$_2$Ir$_2$O$_7$,\cite{kumar} and whether this effect is a result of comparatively higher $U$ in 3$d$ Ti, needs to be understood. Generally, Ti$^{4+}$ with its 3$d^0$ character causes site dilution in magnetic structure, in addition to tuning of both SOC and $U$ parameters. Interestingly, previous studies show magnetism survives to higher concentration of Ti in various iridate systems with different crystal structures. For instance, in Sr$_2$IrO$_4$ system which has layered and squared Ir structure, magnetism exists at least up to 40 \% of Ti substitution.\cite{Gatimu,ge} In honeycomb based Na$_2$IrO$_3$, trace of magnetism has been observed above $\sim$ 26\% of Ti doping where the systems exhibit spin-glass like behavior in higher doped systems.\cite{Manni} The present Y$_2$Ir$_2$O$_7$ has different pyrochlore structure but similar evolution of magnetic behavior emphasizes that both theoretical calculations as well as microscopic experimental tools need to be extended to comprehend this evolution of magnetic and electronic behavior.

\section{Conclusion}
In summary, we have investigated the evolution of the structural, magnetic and transport properties in pyrochlore iridate Y$_2$(Ir$_{1-x}$Ti$_{x}$)$_2$O$_7$. The system shows no structure phase transition with Ti doping, but the structural parameters show systematic changes. Magnetization data revels that magnetic transition temperature is weakly influenced by Ti substitution but the magnetic moment and magnetic frustration decreases with Ti doping. The parent Y$_2$Ir$_2$O$_7$ is in nonequilibrium magnetic state at low temperature showing magnetic relaxation and aging effect, however, the magnetic relaxation rate increases with Ti doping. Raman spectroscopy measurements indicate no changes in structural symmetry but the position as well as linewidth related to A$_{1g}$ mode show an anomaly around the magnetic transition temperatures. Temperature dependent structural investigation using XRD also imply no change of structural symmetry down to low temperature, although structural parameters show unusual changes around magnetic transitions. Resistivity data of Y$_2$(Ir$_{1-x}$Ti$_{x}$)$_2$O$_7$ show insulating behavior throughout the temperature range, however, resistivity is decreased with Ti doping and the electronic transport mechanism follows power law behavior with temperature. Whole series of samples exhibit a negative MR at low temperature which is considered to be a signature of weak localization (WL) effect in SOC dominated system. A crossover from negative to positive MR with temperature has been observed and is understood to arise as an interplay between magnetic moment and magnetic field.

The 3$d$ based non-magnetic Ti$^{4+}$ is substituted in place of 5$d$ Ir$^{4+}$ to tune the spin-orbit coupling (SOC) and electronic correlation ($U$) in Y$_2$Ir$_2$O$_7$. Unlike to magnetic substitution with 4$d$ based Ru$^{4+}$,\cite{kumar} the Ti$^{4+}$ appears to have weaker effect on the magnetic transition temperature. This is probably due to 3$d$ and 4$d$ character of dopants as it has different influence on SOC and $U$ parameters. This Ti$^{4+}$, in addition to tuning both these parameters, also act for site dilution, hence its contribution to magnetic and transport properties will have added effect. For instance, nonmagnetic Ti$^{4+}$ sitting at vertices of Ir tetrahedra will dilute the magnetic interaction. It has been theoretically discussed that SOC induced Dzyaloshinskii-Moriya interaction reduces magnetic frustration in pyrochlore lattice and renders long-range type ordering.\cite{elhajal} Similarly, onset of AIAO-type AFM spin ordering is predicted in pyrochlore structure with high value of $U$.\cite{shinaoka,Ishii,Hon} Faster magnetic relaxation in Figure 7 is an indicative of tuning of SOC. Further, crossover from negative to positive MR with temperature is understood to be due an interplay between SOC and magnetic moment.

\section{Acknowledgment} 
We acknowledge UGC-DAE CSR, Indore for the magnetization, electrical transport and Raman spectroscopy measurements. We are thankful to Dr. Alok Banerjee, Dr. Rajeev Rawat and Dr. Vasant Sathe for the magnetization, resistivity and Raman measurements and discussions, and to Mr. Kranti Kumar, Mr. Sachin Kumar and Mr. Ajay Kumar Rathore for the helps in measurements. We are thankful to AIRF, JNU for low temperature XRD measurements and Mr. Manoj Pratap Singh for the help in measurements. We acknowledge DST-PURSE for the funding. HK acknowledges UGC, India for BSR fellowship.

\end{document}